\documentclass[10pt]{IEEEtran}
\usepackage{caption}
\IEEEoverridecommandlockouts
\usepackage{cite}
\usepackage{subcaption}
\usepackage{mathtools}
\usepackage{multirow}
 \usepackage{enumitem}
\usepackage{pgfplots}
\pgfplotsset{compat=1.18}
\usepackage[ruled,linesnumbered]{algorithm2e}
\usepackage{tikz}

\usepackage{array}
\usepackage{amssymb}

\usepackage{lipsum} 
\usepackage{tikz,xcolor,hyperref}
\definecolor{lime}{HTML}{A6CE39}
\DeclareRobustCommand{\orcidicon}{%
	\begin{tikzpicture}
	\draw[lime, fill=lime] (0,0) 
	circle [radius=0.16] 
	node[white] {{\fontfamily{qag}\selectfont \tiny ID}};
	\draw[white, fill=white] (-0.0625,0.095) 
	circle [radius=0.007];
	\end{tikzpicture}
	\hspace{-2mm}
}
\usepackage{enumitem}

\foreach \x in {A, ..., Z}{%
	\expandafter\xdef\csname orcid\x\endcsname{\noexpand\href{https://orcid.org/\csname orcidauthor\x\endcsname}{\noexpand\orcidicon}}
}\usepackage{ulem}

\usepackage{booktabs}
\usepackage{siunitx}
\usepackage{tikz}
\usetikzlibrary{positioning,arrows.meta,fit,calc,shapes.misc}
\definecolor{a1}{RGB}{230,120,20}    % A1 policy
\definecolor{e2}{RGB}{0,140,90}      % E2 near-RT KPIs
\definecolor{o1}{RGB}{110,70,170}    % O1 / M-Plane (NETCONF/SSH/TLS)
\definecolor{priv}{RGB}{85,85,85}    % operator private API
\definecolor{macsec}{RGB}{0,100,200} % MACsec link
\definecolor{ipsec}{RGB}{10,90,160}  % IPsec tunnel

\begin{document}

\title{\huge Solving the Post-Quantum Control Plane Bottleneck: Energy-Aware Cryptographic Scheduling in Open RAN}

\author{Neha Gupta\orcidA{}~\IEEEmembership{(Student Member,~IEEE)}, Hamed Alimohammadi\orcidB{}, Mohammad Shojafar\orcidC{}~\IEEEmembership{(Senior Member,~IEEE)}, De Mi\orcidD{}~\IEEEmembership{(Senior Member,~IEEE)}, and Muhammad N.M. Bhutta\orcidE{}~\IEEEmembership{(Member,~IEEE)}
\thanks{N. Gupta, H. Alimohammadi, M. Shojafar are with 5G \& 6GIC, University of Surrey, Guildford, UK. De Mi is with the Birmingham City University, United Kingdom, and M.N.M. Bhutta is with Abu Dhabi University, UAE. %email: \{zhizhou.he, m.shojafar, r.tafazolli\}@surrey.ac.uk}
}}

\maketitle

\begin{abstract}

The Open Radio Access Network (O-RAN) offers flexibility and innovation but introduces unique security vulnerabilities, particularly from cryptographically relevant quantum computers. While Post-Quantum Cryptography (PQC) is the primary scalable defence, its computationally intensive handshakes create a significant bottleneck for the RAN control plane, posing sustainability challenges. This paper proposes an energy-aware framework to solve this PQC bottleneck, ensuring quantum resilience without sacrificing operational energy efficiency. The system employs an O-RAN aligned split: a Crypto Policy rApp residing in the Non-Real-Time (Non-RT) RIC defines the strategic security envelope (including PQC suites), while a Security Operations Scheduling (SOS) xApp in the Near-RT RIC converts these into tactical timing and placement intents. Cryptographic enforcement remains at standards-compliant endpoints: the Open Fronthaul utilizes Media Access Control Security (MACsec) at the O-DU/O-RU, while the xhaul (midhaul and backhaul) utilizes IP Security (IPsec) at tunnel terminators. The SOS xApp reduces PQC overhead by batching non-urgent handshakes, prioritizing session resumption, and selecting parameters that meet slice SLAs while minimizing joules per secure connection. We evaluate the architecture via a Discrete-Event Simulation (DES) using 3GPP-aligned traffic profiles and verified hardware benchmarks from literature. Results show that intelligent scheduling can reduce per-handshake energy by approximately 60 percent without violating slice latency targets.

%In a 24-hour emulation with 3GPP-like load and published PQC/accelerator costs, we show that policy-driven batching, placement and offload can reduce per-handshake energy by ~60\%\ without violating slice latency targets.
\end{abstract}

\section{Introduction}
\IEEEPARstart{T}{he} Open Radio Access Network (O-RAN) architecture is built on open, standardized interfaces and disaggregated components, with RAN Intelligent Controllers (RICs) at its core. The Near-Real-Time (Near-RT) RIC (hosting xApps) operates at 10ms–1s timescales for latency-sensitive control, while the Non-Real-Time (Non-RT) RIC (hosting rApps) manages higher-level, longer-timescale optimizations. These controllers enable closed-loop, data-driven optimization of network functions, including security and resource allocation, by leveraging Artificial Intelligence and Machine Learning (AI/ML). As quantum computing threatens traditional cryptography, integrating Post-Quantum Cryptography (PQC) into O-RAN is critical for future-proofing network security.

Data protected today may be recorded and decrypted later. This creates a "harvest-now, decrypt-later" risk: adversaries can capture ciphertext today and decrypt it once large-scale quantum computers exist. Long-lived assets in the RAN, including control-plane sessions, management credentials, policy artifacts, and user data with extended confidentiality requirements need protection that remains robust even decades after deployment. The National Institute of Standards and Technology (NIST) has finalized post-quantum standards, including ML-KEM (Kyber), ML-DSA (Dilithium), and SLH-DSA (SPHINCS+), with guidance emphasizing crypto-agility and hybrid modes during transition~\cite{nist_fips204}. In mobile networks, however, post-quantum security hardening translates into significant operational costs: larger handshakes and keys, more CPU cycles and memory pressure, and increased control-plane traffic. These factors create potential latency exposure for slices with tight budgets, such as Ultra-Reliable Low-Latency Communication (URLLC), putting both latency budgets and capacity under pressure. In this context, we define sustainability as operational energy efficiency, measured in joules per secure connection, ensuring that quantum-safe hardening does not compromise the network's long-term energy-saving targets.

Table~\ref{tab:sustain} summarizes the main sustainability challenges, from session churn and accelerator overheads to geo-temporal carbon variability and telemetry needs. These insights motivate our design choices for embedding security mechanisms into O-RAN. We utilize verified physical benchmarks to demonstrate comparative performance trends and validate the proposed control mechanisms; while absolute energy metrics are implementation-dependent, the relative benefits of protocol-level orchestration and the SOS scheduling logic remain hardware-agnostic and robust across heterogeneous vendor platforms. We embed security operations into O-RAN's existing control loops: a Crypto Policy rApp sets the strategic policy envelope (allowed suites and rekey bounds), and a Security Operations Scheduling (SOS) xApp turns that policy into tactical Near-RT actions. The SOS xApp reduces the impact of computationally expensive PQC handshakes by batching non-urgent work, prioritizing session resumption, steering heavy operations to accelerators, and selecting parameters that meet slice Service Level Agreements (SLAs). This design delivers quantum-resilient security without sacrificing the timing discipline that O-RAN demands.

\textbf{Contributions}
This article makes the following contributions toward an energy-efficient, quantum-resilient O-RAN:
\begin{itemize}[leftmargin=*] \item \textbf{Energy-aware, post-quantum resilient framework.} We design an architecture that embeds PQC control into existing network loops. A Crypto Policy rApp in the Non-Real-Time (Non-RT) RIC defines the strategic policy envelope (allowed suites and rekey bounds), while a Security Operations Scheduling (SOS) xApp in the Near-Real-Time (Near-RT) RIC converts this policy into tactical timing and placement intents.

 \item \textbf{Core mechanisms that mitigate the PQC bottleneck.} We formalize 
three tightly coupled mechanisms: handshake scheduling that nudges non-urgent 
rekeys into low-load windows, session resumption with mobility-aware pre-seeding, and adaptive suite/accelerator selection based on predicted energy gain.

\item \textbf{Concrete decision method.} We specify a 
constrained reinforcement learning (RL) policy in the SOS xApp that minimizes joules per secure 
connection while adhering to hard safety constraints, such as 95th 
percentile (p95) latency targets and minimum security levels. The 
observation vector combines real-time E2SM-KPM signals with an energy 
proxy. An energy proxy is an indirect measurement used to estimate 
joules per operation—for example, CPU cycle counts converted to energy 
using published per-instruction profiles, or a weighted combination of 
accelerator queue depth and utilization metrics.

\item \textbf{Standardized enforcement boundary following Zero Trust principles.} We make explicit that cryptographic control remains at standards-compliant endpoints to maintain O-RAN architectural integrity: the Open Fronthaul utilizes Media Access Control Security (MACsec) at the O-DU/O-RU, while the xhaul (midhaul and backhaul) utilizes IP Security (IPsec) at dedicated tunnel terminators. This ensures the RIC remains an optimization engine for scheduling and placement rather than a functional bottleneck in the data path. \end{itemize}

% ---------- Sustainability challenge table ----------
\begin{table}[t]
\small
\renewcommand{\arraystretch}{1.08}
\caption{Sustainability challenges for post-quantum security in mobile networks.}
\label{tab:sustain}
\centering
\setlength{\tabcolsep}{3pt}
\begin{tabular}{p{0.22\columnwidth} p{0.53\columnwidth} p{0.08\columnwidth}}
\hline
\textbf{Challenge} & \textbf{Impact on Security Operations} & \textbf{Refs} \\
\hline
PQC handshakes \& session churn & More CPU/memory/bytes per handshake; short-lived flows increase handshake frequency and energy per user if unmanaged & \cite{nist_fips204,polese2023understanding} \\
\hline
Accelerators \& carbon variability & Gains must outweigh device/idle power and cooling; identical operations have different carbon footprints by site/time & \cite{kim2020smartnic,wiesner2021wait} \\
\hline
Telemetry \& batching guardrails & Need fine-grained counters to attribute energy; strict per-slice caps required (URLLC never batched) & \cite{bonati2021intelligence,alavirad2023oran} \\
\hline
\end{tabular}
\end{table}

\section{Related Work}

The integration of PQC into O-RAN is motivated by the vulnerability of classical cryptography to quantum attacks. While PQC algorithms (Kyber, Dilithium, Falcon, SPHINCS+) can be incorporated into O-RAN's security protocols (IPsec, MACsec, TLS) without significant performance degradation, the choice of PQC parameters impacts both security and operational efficiency. The O-RAN Alliance recognizes that adopting PQC is challenging due to higher CPU/memory usage and energy draw, calling for an optimal and efficient method to identify the appropriate PQC algorithms for [each] Network Function and interface~\cite{oran2023quantum}.

\textbf{Cipher suite selection and energy profiling}: 
Recent work shows that algorithm choice critically impacts feasibility: SPHINCS+ with HQC KEM incurs prohibitive latency for time-sensitive 5G scenarios, while lattice-based schemes (ML-KEM + Dilithium) are efficient for low-latency use cases~\cite{hoque2025pqc}. Energy profiling on embedded hardware confirms that PQC key establishment is significantly more power-hungry than classical Elliptic Curve Diffie-Hellman (ECDH), with ML-KEM consuming substantially more energy than traditional key exchange on ARM Cortex-M4 platforms~\cite{Patterson2025EnergyPQC}. Detailed energy measurements of post-quantum TLS 1.3 on resource-constrained devices reveal that full mutual authentication handshakes with NIST-standardized algorithms impose measurable energy overhead~\cite{tasopoulos2023energy}, while session resumption using TLS 1.3 PSK mechanisms can reduce energy consumption to approximately 5\% of a full handshake~\cite{restuccia2020performance}. This motivates active scheduling at edge endpoints (O-DU/O-RU, IPsec gateways). AI-driven frameworks using reinforcement learning for adaptive PQC selection have demonstrated 38\% energy efficiency gains and 27\% latency reduction on mobile platforms~\cite{ali2025ai}. However, no prior work implements such intelligent control within O-RAN's RIC architecture.

\textbf{O-RAN architecture and energy orchestration}: 
The O-RAN Alliance defines the SMO/Non-RT/Near-RT split and standard interfaces (A1, E2, O1)~\cite{polese2023understanding,oran-arch}, with E2SM-KPM and E2SM-RC service models enabling xApp control. Prior xApps focus on radio resource management; in contrast, we focus on cryptographic operations scheduling. NIST's finalized standards (ML-KEM, ML-DSA, SLH-DSA)~\cite{nist_fips204} inform cipher choices, while empirical studies confirm PQ-TLS has higher handshake costs than classical TLS~\cite{sosnowski2023pqtls}, motivating selective accelerator offload~\cite{kim2020smartnic}. Energy-aware orchestration demonstrates that temporal workload shifting can cut emissions without violating SLAs~\cite{wiesner2021wait}.

\textbf{Novelty}: To our knowledge, no prior work integrates an O-RAN-aligned xApp/rApp split with PQC-aware security, energy-aware scheduling, and batching/accelerator techniques into a unified framework for sustainable security operations.

 \section{Proposed Energy-Aware Security Orchestration}
The proposed framework leverages the disaggregated control architecture of Open RAN to embed post-quantum security adaptively and energy-efficiently. This hierarchical approach separates strategic policy management from tactical execution, ensuring the transition to PQC does not violate the network's timing discipline.

\subsection{Architecture and Interface Boundaries}
\begin{figure*}[t]
    \centering
    \includegraphics[width=\textwidth]{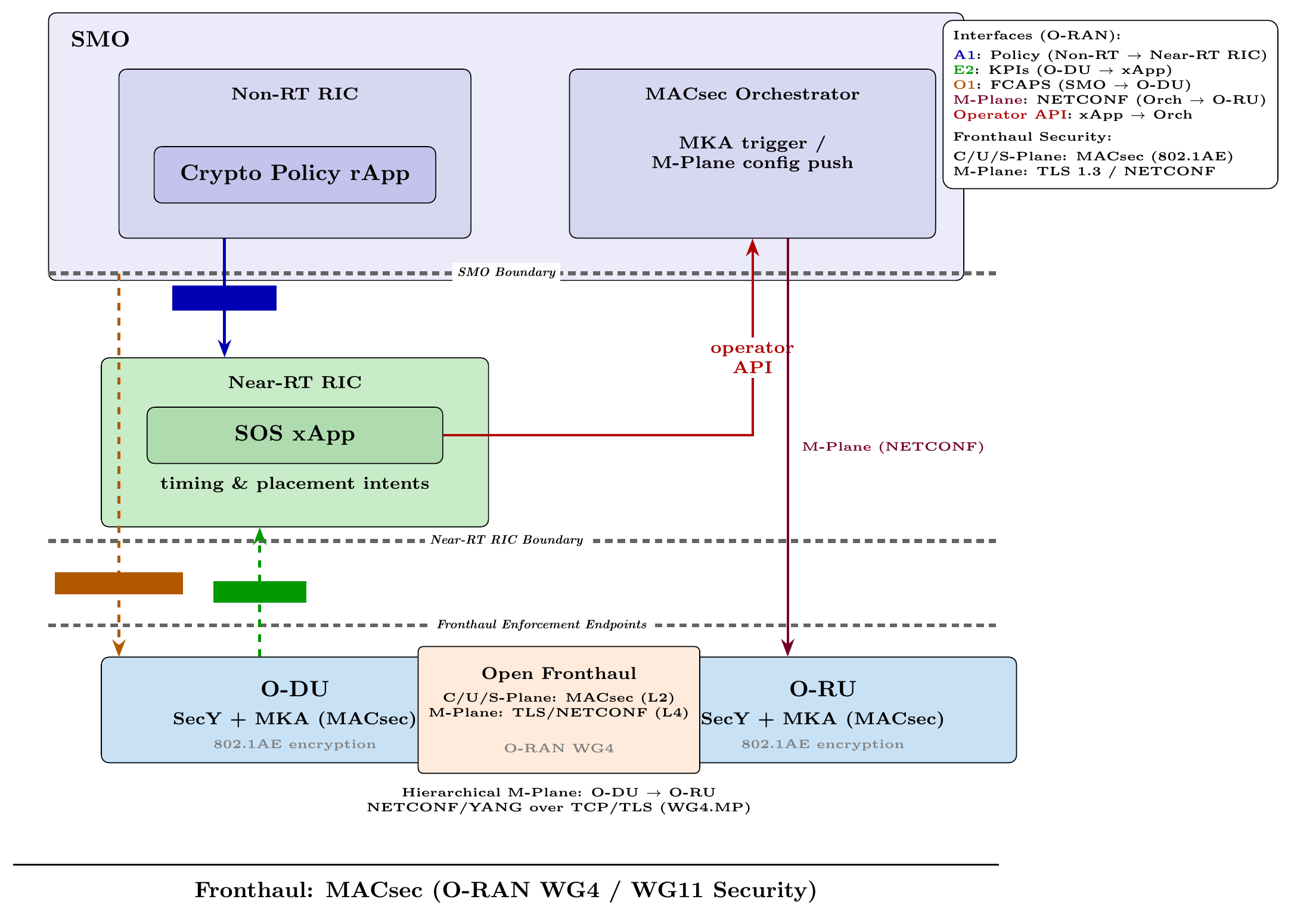}
    \caption{Fronthaul security architecture. The SOS xApp issues timing and placement intents via standard O-RAN interfaces: A1 (policy), E2 (KPI telemetry), and O1 (configuration). Data-plane C/U/S traffic between O-DU and O-RU is secured with MACsec (IEEE 802.1AE); M-plane NETCONF is protected via TLS 1.3. The O-DU manages O-RU settings through hierarchical M-plane control per O-RAN WG4.MP.}
    \label{fig:fronthaul}
\end{figure*}

\begin{figure*}[t]
    \centering
    \includegraphics[width=\textwidth]{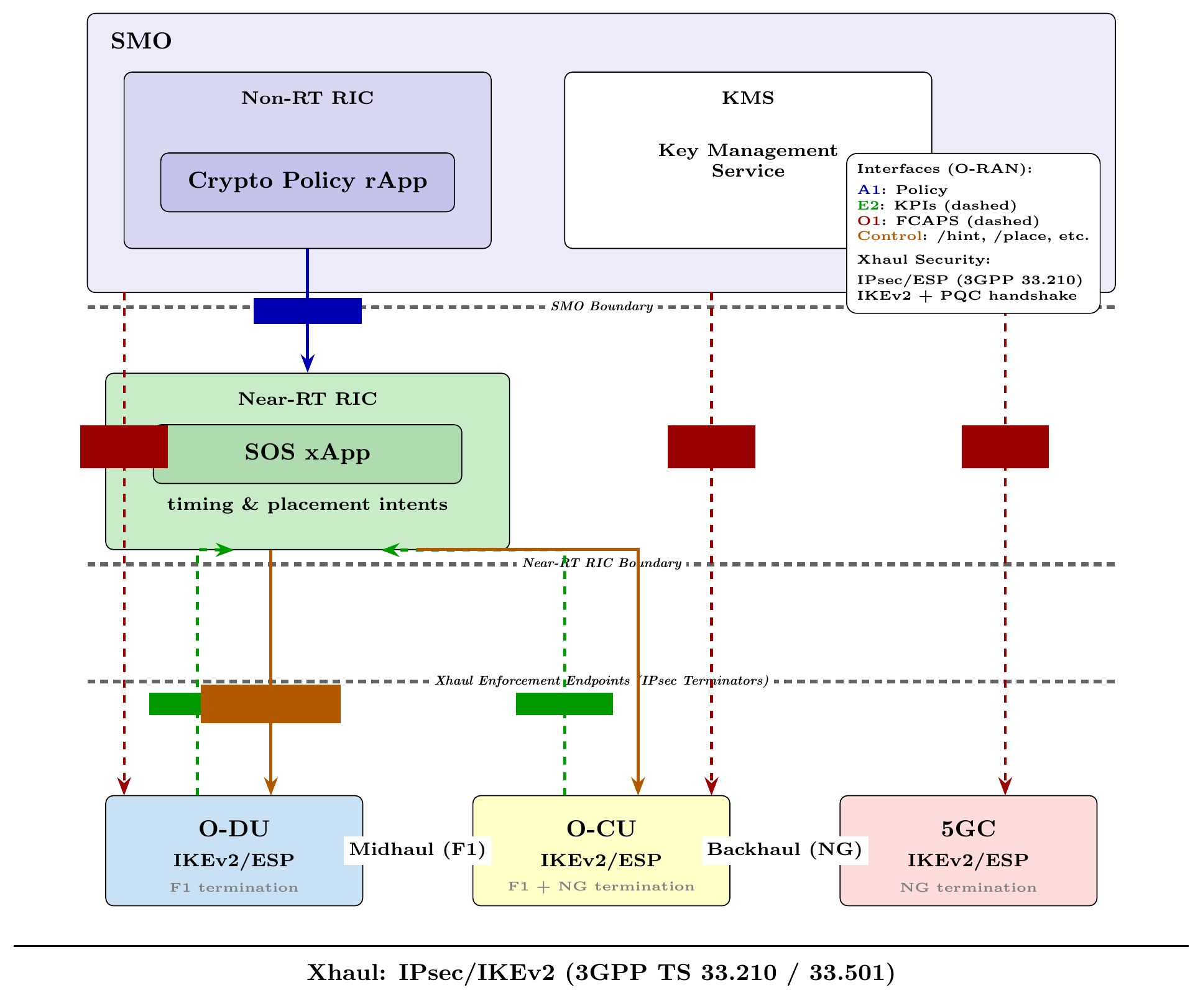}  
    \caption{%
        \textbf{Backhaul/midhaul enforcement architecture.}
        The SOS xApp issues timing and placement intents via O-RAN interfaces: A1 (policy), E2 (KPI telemetry), and O1 (configuration). Control actions (\texttt{/hint}, \texttt{/place}, \texttt{/suite}, \texttt{/offload}) are translated to IKEv2/Encapsulating Security Payload (ESP) parameters at IPsec terminators. Midhaul (F1) secures O-DU$\leftrightarrow$O-CU; backhaul (NG) secures O-CU$\leftrightarrow$5GC. All cryptographic operations remain at the endpoints per 3GPP TS~33.210/33.501, preserving O-RAN architectural boundaries.
    }
    \label{fig:xhaul}
\end{figure*}

The framework shown in  Figure~\ref{fig:fronthaul} and Figure~\ref{fig:xhaul} follows a standard-aligned split between two controllers. A Crypto Policy rApp residing in the Non-RT RIC sets the strategic security envelope specifying allowed PQC suites, rekey intervals, and strategic scheduling guidance and publishes this guidance over the A1 interface. A tactical Security Operations Scheduling (SOS) xApp in the Near-RT RIC then translates these policies into timing and placement decisions every 100 to 500 milliseconds. This control cycle ensures the RIC captures multiple load windows to optimize Enhanced Mobile Broadband (eMBB) batching without interfering with the tighter timing requirements of latency-critical traffic. We utilize an xApp rather than a management-plane function to keep the closed loop close to real-time RAN telemetry and ensure vendor independence via standard E2 and A1 boundaries.

To maintain O-RAN architectural integrity, cryptographic enforcement remains at standards-compliant endpoints. On the Open Fronthaul (O-DU to O-RU)Figure~\ref{fig:fronthaul}, intents are realized at the Media Access Control Security (MACsec) layer via standards-compliant key agreement rekeys or M-Plane provisioning. On the xhaul (midhaul and backhaul)Figure~\ref{fig:xhaul}, intents are executed at the devices that terminate the IPsec tunnels using native management and IKEv2 protocols. The SOS xApp issues timing intents via a narrow E2SM-RC profile, which act as hints for endpoint security agents, while placement intents (site selection) are mediated by the Service Management and Orchestration (SMO) framework over the O1 or M-Plane (NETCONF/YANG) interfaces. We assume an operator-vetted xApp to ensure that decision-making remains trusted and follows Zero Trust principles.

% ==============================================================================
% FIGURE
% ==============================================================================

\subsection{Core Scheduling Mechanisms}
The SOS xApp mitigates the PQC control plane bottleneck through several interlinked mechanisms that optimize the timing and placement of cryptographic workloads.

\textit{Scheduling and optimization Objectives}: The framework balances three goals using a hierarchical satisfy-then-optimize strategy. First, it ensures strict compliance with slice Service Level Agreements (SLAs). While Ultra-Reliable Low-Latency Communication (URLLC) traffic is explicitly exempt from any scheduling delays, Enhanced Mobile Broadband (eMBB) operations are grouped only when the controller detects sufficient latency headroom. Second, the framework minimizes energy by preferring timing intents that create brief batching windows and prioritizing session resumption over full handshakes. Third, it maintains network stability by tracking handshake failure rates and 95th percentile latency; if these metrics regress, the xApp immediately reverts to safe default profiles to prioritize connectivity over efficiency.

\textit{Handshake Batching and Placement}: Computationally expensive PQC work is time-shifted to avoid signaling bursts that could overwhelm the processor. The SOS xApp issues short timing hints that allow handshakes to be executed in tight windows of typically 50 milliseconds during periods of low radio load. On the Open Fronthaul, this intent is realized at the Radio Unit as a scheduled Media Access Control Security (MACsec) rekey, utilizing hitless Secure Association Key (SAK) rollover to prevent packet loss. On the xhaul, the hint facilitates the batching of Internet Key Exchange version 2  (IKEv2) child security association rekeys at the IPsec tunnel terminators. In parallel, the placement intent specifies the optimal enforcement locus such as choosing between an edge node, a regional cloud node, or a dedicated hardware accelerator to minimize the joules consumed per handshake while respecting the slice budget.

\textit{Session Resumption and Lifetime Management}: When the strategic policy defines a rekeying interval, the SOS xApp intelligently schedules these events during quiet intervals to avoid synchronization bursts. On the xhaul and management links, the framework reduces full PQC exchanges by maximizing the use of TLS~1.3 and IKEv2 session resumption. A key functional innovation is the pre-seeding of resumption material at likely target terminators ahead of mobility events, which prevents the need for computationally expensive fresh handshakes during handovers. To further harden the design, rekey timers are frozen during handover storms or high-mobility events unless a safety threshold is reached, yielding lower handshake counts and predictable near-real-time timing without breaching slice budgets.

\textit{Adaptive Suite and Accelerator Policies}: The Non-Real-Time (Non-RT) RIC defines the strategic allow-lists for NIST-standardized algorithms and MACsec policy bounds. Within this envelope, the SOS xApp selects per-slice parameters, such as the security level for ML-KEM, the signature profile, or the use of hybrid modes, based on real-time resource availability. The xApp requests hardware acceleration only when the predicted end-to-end latency with offload remains within the slice budget and when the energy model predicts a net gain over software execution. Strict admission controls and queue-depth caps prevent accelerator contention from inflating tail latency; if performance indicators degrade, the controller seamlessly falls back to software-based execution.

% ==============================================================================
% FIGURE
% ==============================================================================

\subsection{Decision Logic and Runtime Constraints}
The SOS xApp utilizes a constrained reinforcement learning policy to optimize the energy-latency trade-off. Each control cycle, the agent observes an observation vector summarizing slice latency headroom, cell load, mobility rates, and an energy proxy. 

\textit{Sustainability Metrics and timing}: To respect the Near-RT RIC timing constraints (10~ms to 1~s), the SOS xApp utilizes a characterization-based abstraction. Energy consumption for PQC operations such as the 17.57~mJ required for a full mutual handshake or the 0.88~mJ for a session resumption is characterized offline through platform profiling and stored as static cost coefficients~\cite{tasopoulos2023energy}. The energy proxy thus represents a pre-configured mapping of these joules-per-handshake values to the real-time operational context. At runtime, the xApp observes session state (e.g., RRC-Connected vs. Idle) and handover events via standard E2SM-KPM and E2SM-RC counters. This separation ensures that scheduling decisions meet RIC latency requirements without being blocked by the relative sparsity of standardized hardware power reports or the high latency of carbon-grid metrics. Carbon-aware scheduling—relying on signals with update intervals of 15 minutes or more is appropriately deferred to the Non-RT RIC layer.

\textit{Control Cycle}: The SOS xApp operates on a 100--500~ms control cycle, selected based on three specific constraints. A lower bound of 100~ms is established because a full PQC handshake requires approximately 98.48~ms of computation time; control decisions faster than this interval cannot observe handshake completion and would provide no additional benefit. An upper bound of 500~ms ensures the controller issues at least one tactical scheduling decision before potential SLA violations for standardized eMBB traffic (300~ms delay budget). This duration is carefully aligned with O-RAN WG4 specifications for fronthaul time-budgets where transport hop delays (modeled via T12 and T34 timers) are typically limited to 250 microseconds ensuring that handshakes can be batched without exceeding the 150 millisecond broadband SLA.This range adequately captures eMBB load windows (typically 50--100~ms), allowing the xApp to identify bursty traffic gaps and nudge resumed handshakes into gaps without competing with data traffic. Crucially, this cycle ensures URLLC non-interference, as traffic with strict 10~ms budgets operates on timescales much faster than the xApp control cycle; these flows bypass scheduling entirely to maintain their strict latency requirements.

\subsection{SLA-Sustainability Conflict Resolution} The SOS xApp resolves conflicts between sustainability objectives and SLA constraints through a strict priority hierarchy that ensures network reliability is never sacrificed for efficiency. This safety-centric design follows a satisfy-then-optimize strategy.

\textit{Traffic Prioritizing and Guarding}: The framework establishes absolute priority for latency-critical traffic classes, such as Ultra-Reliable Low-Latency Communication (URLLC). These flows are never subject to scheduling delays or batching windows; the xApp selects the efficient session resumption path only if a valid security context exists and can be executed immediately. For broadband traffic, the xApp performs a continuous headroom check. It estimates total latency as the sum of current queue residency and expected service time; if this estimate approaches a predefined safety margin of the SLA budget, the controller disables energy-driven deferral and processes the request immediately.

\textit{Borderline Cases and Stressed Conditions}: To maintain robustness under stressed network conditions, the scheduling logic incorporates a functional safety shield. During mobility bursts such as handover storms triggered by high-speed mobility where session continuity degrades the SOS xApp detects the elevated frequency of mobility events via standard E2SM-KPM execution counters. In response, the safety shield automatically relaxes energy optimization targets and prioritizes session stability to prevent signaling congestion. Similarly, if queue residency exceeds a critical threshold (indicating accelerator contention), the framework enters a safety mode that prioritizes throughput over efficiency. In this state, batching and deferral are suspended until the congestion clears. This hierarchical approach ensures that the RIC remains an optimization engine that respects the fundamental timing discipline of the Radio Access Network.

\section{Case Study: Energy-Aware Security Scheduling}
\label{sec:case_study}

We evaluate the SOS framework through discrete-event simulation to quantify the energy-latency trade-off of PQC handshake scheduling in O-RAN deployments. Parameters are derived from peer-reviewed measurements and 3GPP 
specifications where available and the model assumptions are explicitly stated.

\subsection{Simulation Methodology}

We model an urban macro-cell deployment with 100 O-RUs over a 24-hour operational cycle, generating approximately 216,000 security events. The arrival process follows a Poisson distribution with a mean rate of 90 handshake requests per hour per O-RU, consistent with the urban vehicular handover frequency at 60 km/h specified in 3GPP TR 38.913 Section 7.7.

Each security event is classified as either a full PQC handshake or a resumed handshake based on session state. For full handshakes using ML-KEM-768 and ML-DSA-65, we use 17.57 mJ energy and 98.48 ms computation time, drawn from Tasopoulos et al.~\cite{tasopoulos2023energy} who profiled NIST-standardized PQC algorithms on ARM Cortex-M4 platforms at 180 MHz. For resumed handshakes using TLS 1.3 Pre-Shared Key(PSK), we model energy and time as five percent of full handshake values (0.88 mJ, 4.92 ms). This ratio is a conservative estimate motivated by Restuccia et al., who found that classical PSK-based exchanges consume approximately 2.5 percent of ECDHE-ECDSA energy; we double this to account for PQC symmetric overhead~\cite{restuccia2020performance}. The probability of successful session resumption is inversely related to mobility, as higher handover rates increase session breaks and necessitate more full handshakes. We evaluate three scenarios: Baseline with zero percent resumption representing legacy behavior where every event triggers full authentication, SOS-Low with 40 percent resumption typical of highway mobility at 120 km/h, and SOS-High with 63 percent resumption achievable in urban vehicular conditions. The SOS xApp enforces a 150 ms latency threshold for security operations, derived from the eMBB packet delay budget in 3GPP TS 23.501.

\subsection{Results}

\begin{figure}[t]
\centering
\includegraphics[width=\columnwidth]{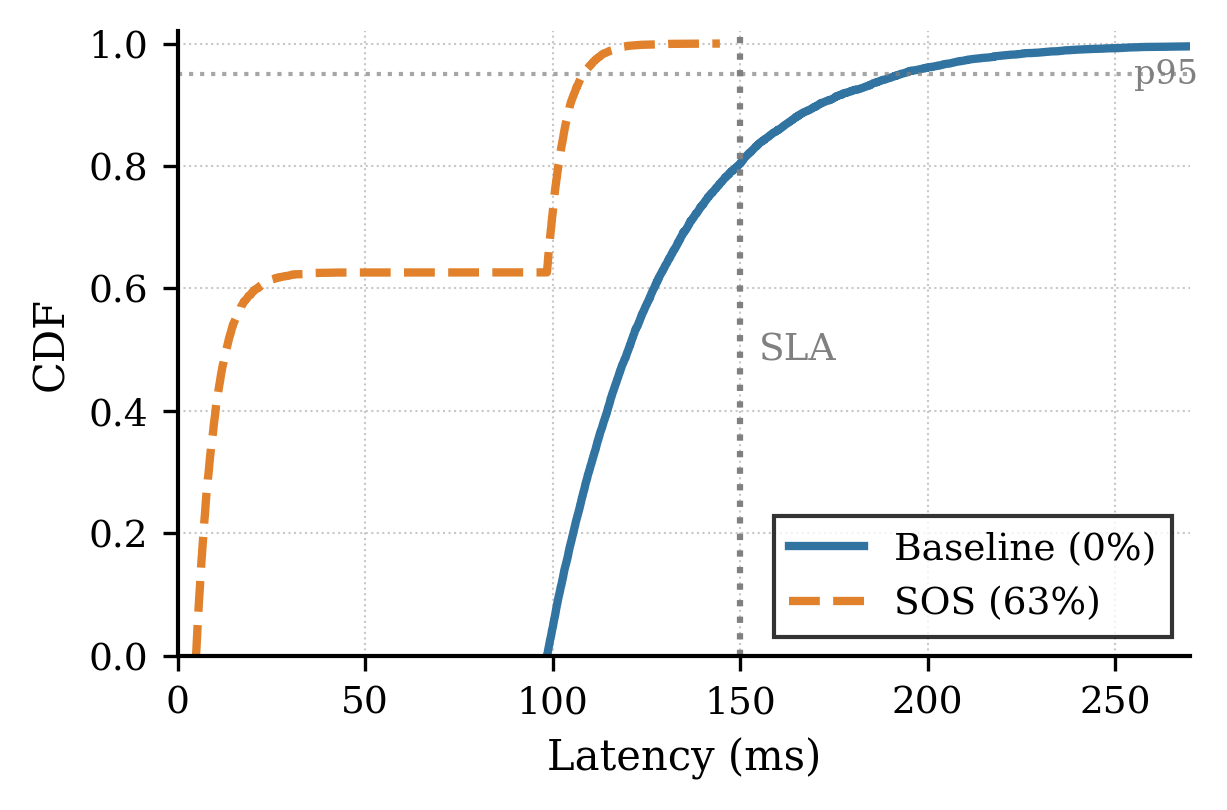}
\caption{Handshake latency CDF. Baseline (blue solid): all full PQC handshakes with 98.48 ms service time. SOS (orange dashed): 63 percent session resumption with 4.92 ms resumed handshake time. The vertical dotted line indicates the 150 ms SLA threshold; horizontal dotted line marks the 95th percentile.}
\label{fig:latency_cdf}
\end{figure}

Figure~\ref{fig:latency_cdf} presents the cumulative distribution function of handshake latency for baseline and SOS configurations. The baseline curve (blue solid line) exhibits a distribution centered around 100 ms, reflecting the deterministic service time of full PQC handshakes. In contrast, the SOS distribution (orange dashed line) is bimodal: 63 percent of requests complete within approximately 5 ms as resumed handshakes, while the remaining 37 percent follow the full handshake path. This shift substantially reduces tail latency, with the 95th percentile dropping from 191 ms in the baseline to 98 ms with SOS, ensuring compliance with the 150 ms SLA threshold indicated by the vertical dotted line.

\begin{figure}[t]
\centering
\includegraphics[width=\columnwidth]{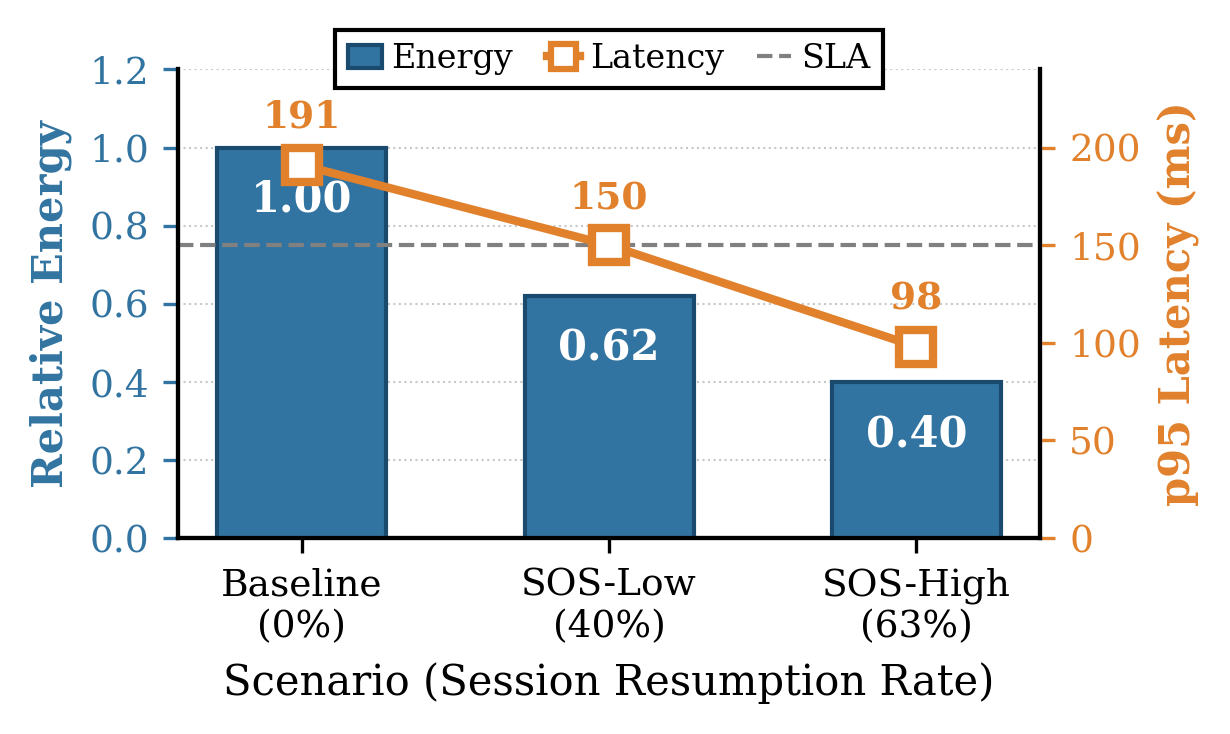}
\caption{Energy-latency trade-off across session resumption scenarios. Bars (left axis, blue): relative energy consumption normalized to baseline. Line with markers (right axis, orange): absolute p95 latency in milliseconds. Dashed horizontal line indicates the 150 ms SLA threshold. As resumption rate increases from zero to 63 percent, energy decreases by 60 percent and p95 latency decreases by 48 percent, demonstrating that both metrics improve in parallel.}
\label{fig:energy_latency}
\end{figure}

Figure~\ref{fig:energy_latency} illustrates the dual benefits of session resumption scheduling using a combined bar-line visualization. The left axis with blue bars shows relative energy consumption normalized to the baseline, while the right axis with orange line and markers shows absolute p95 latency in milliseconds. Three key observations emerge from this figure.

First, energy reduction scales with resumption rate. Relative energy decreases from 1.00 at baseline to 0.62 at SOS-Low with 40 percent resumption to 0.40 at SOS-High with 63 percent resumption, representing up to 60 percent reduction in per-handshake energy consumption.

Second, latency improves in parallel. The p95 latency decreases from 191 ms at baseline to 150 ms at SOS-Low to 98 ms at SOS-High, a 48 percent improvement that maintains comfortable margin below the SLA threshold shown by the dashed horizontal line.

Third, there is no trade-off between objectives. Unlike typical optimization problems where improving one metric degrades another, session resumption simultaneously reduces both energy and latency because resumed handshakes are inherently faster and less computationally intensive than full PQC exchanges.

Under the simulated urban mobility scenario with 63 percent session resumption, the framework achieves 60 percent reduction in energy consumption compared to a baseline where every security event triggers a full PQC handshake. The average energy per handshake decreases from 17.57 mJ in the baseline to 7.06 mJ using the SOS framework. SLA compliance improves from 85.7 percent in the baseline to 98.2 percent with SOS-High, as fewer requests experience the lengthy full handshake processing time.

\subsection{Sensitivity to Mobility}

Energy savings range from 38 percent (highway, 120 km/h) to 85 percent (indoor/stationary), scaling with achievable resumption rates per 3GPP TR 38.913 mobility profiles.

\subsection{Discussion}

The approximately 20-fold difference between full and resumed handshake 
costs reflects TLS 1.3 PSK mechanics, where session resumption bypasses 
expensive asymmetric operations. This ratio persists regardless of whether cryptographic operations execute on embedded processors or dedicated accelerators, making the SOS approach robust across heterogeneous O-RAN deployments.

For operators planning PQC migration, these results suggest that investment in session management infrastructure including PSK caching, session ticket distribution, and intelligent resumption scheduling yields substantial sustainability benefits without requiring hardware upgrades. The SOS xApp provides a software-defined path to energy-efficient post-quantum security that complements rather than replaces hardware acceleration strategies.

% ==============================================================================
% FIGURES
% ==============================================================================

\section{System Implications and Future Directions}

The evaluation demonstrates that protocol-level session management is the 
most impactful lever for control-plane energy optimization in post-quantum 
O-RAN deployments. The approximately 20-fold cost difference between full 
and resumed handshakes is a hardware-agnostic property of TLS 1.3, ensuring 
the SOS scheduling logic delivers consistent relative benefits across 
heterogeneous platforms.

For practical deployment, we recommend a staged rollout: placement-only 
mode first, then adaptive accelerator admission with concurrency caps, 
and finally intelligent batching once per-cell latency slack is characterized.

Several areas remain for future standardization and research. The 
simulation-based evaluation presented here should be validated on O-RAN 
testbed infrastructure as PQC-enabled RIC platforms become available. 
A narrow E2SM-RC profile enabling security-aware scheduling nudges would 
formalize the xApp control interface proposed here. Calibrated hardware 
energy counters for individual cryptographic operations—currently absent 
from O-RAN and 3GPP specifications—would enable more accurate offload 
accounting. Carbon-aware scheduling, requiring grid intensity signals 
with update intervals of 15 minutes or more, is appropriately scoped to 
Non-RT RIC or SMO layers. Integrating the SOS xApp with conflict-aware 
coordination frameworks such as PACIFISTA~\cite{delprever2024pacifista} 
will ensure security-driven decisions do not interfere with power-saving 
rApps. Finally, extending this framework to multi-connectivity and 
non-terrestrial network scenarios in 6G will require meta-learning 
approaches to handle extreme channel variability.

\vspace{-15px}
\section*{Acknowledgment}
This work has been supported by the ORAN-TWIN-X subproject under CHEDDAR: Communications Hub for Empowering Distributed Cloud Computing Applications and Research funded by the UK Engineering and Physical Sciences Research Council (EPSRC) under grant numbers EP/Y037421/1 and EP/X040518/1.
\vspace{-10px}

\bibliographystyle{unsrt}

\bibliography{biblio}

\end{document}